\documentclass{aa}

\usepackage{graphicx}
\usepackage{dcolumn}
\usepackage{bm}
\usepackage{aas_macros}
\usepackage{hyperref}

\usepackage{txfonts}
\usepackage{multirow}

\usepackage{graphicx}
\usepackage{xcolor}
\usepackage{soul}

\begin{document}

\title{Future prospects for measuring 1PPN parameters\\using observations of S2 and S62 at the Galactic Center}
\titlerunning{Measuring 1PPN parameters at the Galactic Center}

\author{Victor de Mora Losada \inst{1}, Riccardo~Della~Monica \inst{2}\fnmsep\thanks{Email: rdellamonica@usal.es}, Ivan de Martino \inst{2,3} and Mariafelicia De Laurentis\inst{4,5}}
\authorrunning{V. de Mora Losada, R.~Della~Monica, I. de Martino, M. de Laurentis}
\institute{Universidad Complutense de Madrid, E-28040 Madrid, Spain
\and Departamento de Física Fundamental, Universidad de Salamanca, Plaza de la Merced, s/n, E-37008 Salamanca, Spain \and Instituto Universitario de Física Fundamental y Matemáticas, Universidad de Salamanca, Plaza de la Merced, s/n, E-37008 Salamanca, Spain \and Universit\'a di Napoli "Federico II", Compl.~Univ.~di Monte S.~Angelo, Edificio G, Via Cinthia, I-80126, Napoli, Italy \and INFN Sezione di Napoli, Compl.~Univ.~di Monte S.~Angelo, Edificio~G, Via Cinthia, I-80126, Napoli, Italy}

\date{Received 30 October 2024 / Accepted 26 January 2025}

\abstract
{The Parameterized Post-Newtonian (PPN) formalism offers an agnostic framework for evaluating theories of gravity that extend beyond General Relativity. Departures from General Relativity are represented by a set of dimensionless parameters that, at the first order in the expansion, reduce to $\beta$ and $\gamma$, which describe deviations in spatial curvature and non-linear superposition effects of gravity, respectively. 
}{We exploit future observations of stars at the Galactic Center, orbiting the supermassive black hole Sagittarius A*, to forecast the ability to constrain the first-order PPN parameters $\gamma$ and $\beta$.}{We have generated a mock catalog of astrometric and spectroscopic data for S2, based on the Schwarzschild metric, simulating observations over multiple orbital periods with the GRAVITY and SINFONI instruments. Our analysis includes the effects of relativistic orbital precession and line-of-sight (LOS) velocity gravitational redshift. 
Since future data for S2 can provide constraints only on a linear combination of 
the PPN parameters $\beta$ and $\gamma$ 
we also analyzed the impact of future observations of 
the gravitational lensing for stars that pass closer in the sky to Sgr A*, like the known star S62, 
which can potentially provide tight constraints on the parameter $\gamma$, that alone regulates the amplitude of the astrometric deviations due to lensing. 

}{When combining lensing observations for S62, and the precise orbital tracking of S2, one obtains independent constraints on both $\gamma$ (with a potential precision as good as $\sim 1\%$) and $\beta$ (with a corresponding precision of $\sim 2\%$), 
providing a 
precision test 
of General Relativity and its extensions.}{}
                    
\maketitle

\section{Introduction}
\label{sec:intro} 

The study of relativistic orbital dynamics around supermassive black holes offers a unique opportunity to test General Relativity and its extensions in a regime inaccessible with classical tests \citep{Psaltis2008}. In recent years, 
thanks to a series of groundbreaking observational campaigns, the Galactic Center of the Milky Way has stood out as a unique and powerful laboratory at the edge of fundamental physics and astrophysics \citep{DeLaurentis2023}. The discovery that Sagittarius A* (Sgr A*), a supermassive compact object, resides at the heart of the Milky Way \citep{Genzel2010}, along with the advent of high-angular resolution instrumentation capable of probing small spatial scales, has driven numerous scientific breakthroughs over the past three decades \citep{Genzel2022}. 
These high-resolution observations have provided new confirmations of classical tests of General Relativity, now in a previously untested 
and extreme environment. The identification of a dense population of rapidly moving stars \citep{Eckart1996, Ghez1998}, known as S-stars, orbiting Sgr A* and observable in the near-infrared with increasingly high precision \citep{Gillessen2017}, has enabled detailed tests of relativistic effects such as the orbital precession \citep{GravityCollaboration2020} and gravitational redshift \citep{, GravityCollaboration2018a} around a supermassive compact object. In addition, direct imaging of Sgr A* and detection of its shadow \citep{EventHorizonTelescopeCollaboration2022a}, using revolutionary techniques such as Very Long Baseline Interferometry, has allowed for the observation of light deflection in a strong gravitational field regime. 
Such high-resolution observations of the Galactic Center offer a unique opportunity not only to test the 
predictions of General Relativity \citep{Will2008, Zucker2006, Merritt2010b}, but also to falsify it against alternative theories 
of gravity \citep{DeLaurentis2023}.

In this context, the Parameterized Post-Newtonian (PPN) formalism provides a powerful framework for comparing General Relativity with alternative theories of gravity, particularly in the weak-field limit \citep{Poisson2014, Will2018}. Introduced in the mid-20th century \citep{Nordtvedt1969, Will1971, Will1972} as a way to systematically classify deviations from General Relativity, the PPN formalism allows different gravitational theories to be tested against one another in a consistent manner \citep{Will2014}. The formalism expands the space-time metric in terms of small perturbations around a Newtonian background, introducing a set of ten dimensionless parameters that quantify potential deviations from the predictions of General Relativity. Among these parameters, $\gamma$ and $\beta$, appearing at first order in the expansion, are of particular importance: $\gamma$ measures how much space curvature is produced by a unit mass, and $\beta$ quantifies the non-linearity of the gravitational superposition law. In General Relativity, both parameters are expected to be exactly equal to one, and any deviation from these values would indicate a departure from Einstein’s theory of gravity. Historically, the first significant test of the PPN formalism, even before its actual formulation, occurred with the measurement of the deflection of light by the Sun during the 1919 solar eclipse \citep{Eddington1920}, which provided the first experimental confirmation of Einstein's predictions. Since then, the formalism has been used in a variety of Solar System experiments \citep{Will2014} to place stringent constraints on $\gamma$ and $\beta$. One of the most precise measurements of $\gamma$ came from the Cassini spacecraft, which measured the Shapiro time delay of radio signals as they passed near the Sun, constraining $\gamma$ to within a few parts in $10^5$ \citep{Bertotti2003}. Similarly, the perihelion precession of Mercury, as well as Lunar Laser Ranging experiments, have provided strong constraints on $\beta$ \citep{Genova2018}. These tests, which probe the weak-field regime, have so far shown excellent agreement with General Relativity, yet the strong-field regime, as present near supermassive black holes, remains relatively unexplored.

This paper builds upon earlier efforts to explore the use of S-star orbits in testing General Relativity and its extensions \citep{DeMartino2021, DellaMonica2022a, DellaMonica2022b, DellaMonica2023a, DellaMonica2023b, Fernandez2023, DeMartino2024}. Here, we systematically analyze the deviations in the astrometric and spectroscopic observables for these objects, within the PPN formalism, by building an orbital model for both massive test particles and photons that includes all the observable relativistic effects (Section \ref{sec:model}). We consider the orbital dynamics of S2 and, through the generation of a mock catalog 
and the application of Markov Chain Monte Carlo (MCMC) techniques (Section \ref{sec:mcmc}), we 
forecast the accuracy 
on the PPN parameters that will arise with observations over multiple periods with the Very Large Telescope (VLT) instruments GRAVITY and SINFONI. The results of this analysis are highlighted in Section \ref{sec:results}. Moreover, we consider the inclusion of the slowly moving star S62, which is supposed to pass in the next few years very close in the sky to Sgr A*, potentially enabling the first direct detection of gravitational lensing for these stars (Section \ref{sec:s62}) 
and investigate its possible impact on constraining the PPN parameters. Finally, we give our final conclusions in Section \ref{sec:discussion}.

\section{The 1PPN orbital model for S2}
\label{sec:model}

The PPN formalism is a theoretical framework that allows the comparison of different gravitational theories in the weak-field, slow-motion regime in which the velocities of physical bodies (both the sources and the test particles) are much smaller than the speed of light \citep{Nordtvedt1969, Will1971, Will1972, Will2018}. It expands from the standard Post-Newtonian (PN) formalism which is a perturbative expansion of the space-time metric that assumes the validity of General Relativity. Instead, the PPN formalism provides a systematic way to express deviations from Newtonian gravity and General Relativity by explicitly introducing dimensionless parameters that quantify these differences. This allowed the PPN formalism to play an essential role in the falsification of General Relativity in astrophysical scenarios where relativistic effects are present but not dominant, such as in the Solar System and for weak-field tests in the surrounding of black holes. 
In its general formulation, the PPN formalism introduces a set of 10 dimensionless parameters \citep{Will2018}, which have a clear limit to General Relativity and encode space curvature, nonlinear gravitational behaviours and preferred-frame effects \citep{Will1972}.

In this work, we assume that the stellar orbits, particularly that of the S2 star, in the Galactic Center satisfy the weak-field, slow-motion assumption and we describe the gravitational field of Sgr A* by means of a static and spherically symmetric solution that is obtained by the PPN expansion of the Schwarzchild metric in General Relativity. When expressed in the standard Schwarzschild coordinates ($t$, $r$, $\theta$, $\phi$), being $r$ the aerial radius, and truncating the expansion at 1PN order, this space-time metric takes the expression
\begin{align}
    ds^2 = g_{tt} dt^2 + g_{rr} dr^2 + r^2 d\Omega^2,
    \label{eq:1PPN_metric}
\end{align}
being 
the solid angle $d\Omega^2 = d\theta^2 + \sin^2\theta d\phi^2$ and the metric coefficients
\begin{align} 
\label{eq:gtt}    g_{tt} &= -1+\frac{2M}{r} + \frac{2M^2(\beta-\gamma)}{r^2},\\
\label{eq:grr}    g_{rr} &= 1+\frac{2M\gamma}{r},
\end{align}
with $M$ the mass of the central object. 

The PPN metric in Eq. \eqref{eq:1PPN_metric} depends on two additional parameters $\gamma$ and $\beta$. Interestingly, these two classical PPN parameters have a physical interpretation, with $\gamma$ encoding how much space curvature a test mass produces, and $\beta$ quantifying the non-linearity in the superposition law for gravity. It is easily shown that the limits $\beta,\gamma\to 1$ reduce Eq. \eqref{eq:1PPN_metric} to the 1PN limit of the Schwarzschild metric, thus reducing the theory to General Relativity. Experimentally, the PPN parameters have been constrained in different scenarios, due to the different manner they impact the relativistic observables. The parameter $\gamma$ is known to directly affect the propagation of light rays in a given metric theory of gravity, for this reason it has been constrained with increasingly high precision via classical tests in the Solar System, including deflection of light and the Shapiro delay for light rays that graze the surface of the Sun. Currently, the most stringent constraint on $\gamma$ has been obtained by the Cassini mission \citep{Bertotti2003}, corresponding to $|\gamma-1|\lesssim2.3\times10^{-5}$. Constraints on the parameter $\beta$, on the other hand, can only be obtained at 1PN order via the study of the orbital motion of test particles, in which both the PPN parameters $\beta$ and $\gamma$ enter, through some linear combination. For this reason, constraints on $\beta$ have been obtained by assuming the Cassini value for $\gamma$, using the measurement of the Nordtvedt effect in Lunar Ranging measurements \citep{Hofmann2010} and from the measurement of the pericenter advance of planets \citep{Will2014}. The currently most stringent limit on $\beta$, $|\beta-1|\lesssim8\times10^{-5}$, comes from the measurement of Mercury's precession by the Messenger spacecraft \citep{Genova2018}.

In this article, we leave both $\gamma$ and $\beta$ as free parameters in the description of the orbits of the S2 star in the Galactic Center, in order to assess the 
accuracy that future observations of this star with the GRAVITY interferometer, spanning two entire orbital periods, will achieve on 
constraining these parameters. In order to do so, we take advantage of the high mass ratio between the star's mass ($\lesssim10M_\odot$) and that of the central black hole ($\sim4\times10^6M_\odot$) to treat the former as a freely falling test particle in the gravitational field of the central object. This also relies on the assumption 
that the perturbation of the orbit from other stars in the cluster and the contribution of an extended mass component are negligible \citep{Lechien2024, Gravity2024}.

Freely falling test particles in the space-time in Eq. \eqref{eq:1PPN_metric} 
move on 
geodesic trajectories, that are described by the following set of equations:
\begin{align}
    \ddot{t} &= \frac{2 M \dot{r} \dot{t} \left(2 M \left(\beta - \gamma\right) - r\right)}{r \left(2 M^{2} \beta - 2 M^{2} \gamma - 2 M r + r^{2}\right)}, \label{eq:geo_t}\\
    \ddot{r} &= \frac{M \gamma r \dot{r}^{2} - M \dot{t}^{2} \left(2 M \left(\gamma- \beta\right) + r\right) + r^{4} \left(\dot{\phi}^{2} \sin^2\theta +  \dot{\theta}^{2}\right)}{r^{2} \left(2 M \gamma + r\right)}, \label{eq:geo_r}\\
    \ddot{\theta} &= \frac{\dot{\phi}^{2} \sin(2 \theta)}{2} - \frac{2 \dot{r}\dot{\theta}}{r},\label{eq:geo_theta} \\   
    \ddot{\phi} &= - \frac{2\dot{\phi}}{r}\left(r\dot{\theta}\cot\theta + \dot{r}\right),\label{eq:geo_phi}
\end{align}
where dots represent derivatives with respect to the particle’s proper time $\tau$. For time-like particles, the normalization condition $g_{\mu\nu}\dot{x}^\mu\dot{x}^{\nu}= -1$ holds. The invariance of this quantity along geodesic trajectories allows us to assign initial conditions that satisfy the normalization, knowing that it holds for the entire evolution. By specifying the initial position $\{t(\tau_0),\,r(\tau_0),\,\theta(\tau_0),\,\phi(\tau_0)\}$ and three, out of the four, components of the 4-velocity, we can determine the fourth via the normalization condition (in this case, we solve for $\dot{t}(0)$ as a function of $\dot{r}(0)$, $\dot{\theta}(0)$, and $\dot{\phi}(0)$). This leaves us with six degrees of freedom that correspond to the initial position and velocity of the particle.

In celestial mechanics, these initial conditions are often expressed through Keplerian orbital elements: the time of passage at pericenter $t_p$, the semi-major axis $a$, the eccentricity $e$, the inclination $i$, the longitude of the ascending node $\Omega$, and the argument of pericenter $\omega$. The first three determine the orbital motion within the orbital plane, while the remaining three define its orientation relative to an observer. In the spherically symmetric case, the symmetry of the system simplifies the dynamics, as orbits confined to the equatorial plane with $\theta = \pi/2$ and $\dot{\theta} = 0$ remain within this plane ($\ddot{\theta} = 0$). Therefore, we can integrate the equations of motion on the equatorial plane and subsequently rotate the trajectory to the observer's reference frame. In Newtonian mechanics, these Keplerian elements uniquely describe elliptical orbits. However, in the PPN framework, deviations from this behavior are expected due to corrections in the gravitational potential (such as terms scaling as $\sim 1/r^3$). These corrections, that are encoded in the geodesic equations Eqs. \eqref{eq:geo_t}-\eqref{eq:geo_phi}, cause the orbital elements to evolve over time. For this reason, the Keplerian elements at a specific time represent an osculating ellipse, approximating the particle's instantaneous trajectory in the PPN formalism. The most noticeable effect on the orbit is a secular precession on the orbital plane, the periapsis advance, which corresponds to
\begin{equation}
    \Delta \omega_\textrm{PPN} = \frac{6\pi GM \Upsilon}{ac^2(1-e^2)}
    \label{eq:PPN_precession}
\end{equation}
per orbital period, where we have introduced the linear combination of PPN parameters
\begin{equation}
    \Upsilon = \frac{2+2\gamma-\beta}{3}.
    \label{eq:upsilon}
\end{equation}
For all values of $\beta$ and $\gamma$ that satisfy the relation $2\gamma-\beta = 1$, corresponding to $\Upsilon = 1$, the general relativistic value of the orbital precession 
is recovered. This feature exhibits an important behavior of massive particles orbits in the PPN formalism, which alone can only help constrain a combination of $\beta$ and $\gamma$. The degeneracy cannot be broken unless the relativistic information from the photons (the lensing, the Shapiro delay, and the gravitational redshift) is included in the model and captured by the data.

In this work, we numerically compute the relativistic sky-projected trajectory of S2 by integrating Eqs. \eqref{eq:geo_t}-\eqref{eq:geo_phi} for a set of orbital parameters. The key steps of this integration procedure are presented here, with more details available in previous works \citep{DeMartino2021, DellaMonica2022a}. The orbital parameters are given by:
\begin{equation}
    \bm{\theta}=\left(D,\,M,\,t_P,\, a,\, e,\, i,\, \Omega,\, \omega,\,\beta,\,\gamma\right),
    \label{eq:parameters}
\end{equation}
where $D$ and $M$ are the distance and the mass of the central object, respectively, while the PPN parameters $\beta$ and $\gamma$ represent our parameters of interest. The remaining elements $(t_p, a, e, i, \Omega, \omega)$ specify the initial conditions for the particle’s trajectory, as mentioned above. Specifically, the parameters $a$ and $e$ define two turning points, the pericenter $r_p = a(1-e)$ and the apocenter $r_a = a(1+e)$, which correspond to the roots of the effective potential. These values fix the initial radial and angular velocities, $\dot{r}(\tau_0)$ and $\dot{\phi}(\tau_0)$. To fully define the trajectory, we set the initial time at the last apocenter passage, $t(\tau_0) = t_a = t_p - T/2 \approx 2010.35$, where $T$ is the orbital period. The geodesic equations are then numerically integrated using an adaptive Runge-Kutta method to match a given observational time span.

\begin{figure}
    \centering
    \includegraphics[width=\columnwidth]{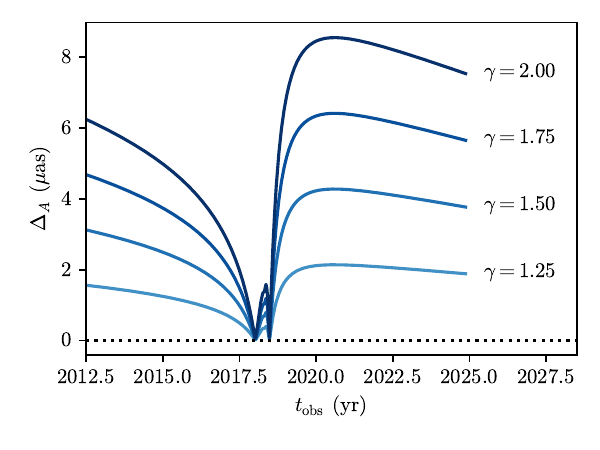}
    \caption{Deviations in the astrometric observable for the S2 star caused by the different photon propagation alone (\textit{i.e.} not considering modifications to the star's orbit) in theories with values of $\gamma$ that differ from the general relativistic case ($\gamma = 1$). A variation of $\gamma$ of order 100\% from the general relativistic case produces an astrometric deviation which is below the instrumental sensitivity of GRAVITY, at 10 $\mu$as.}
    \label{fig:astrometric_lensing}
\end{figure}

The integrated orbits are then converted into the physical quantities that are observed experimentally. These are: \textit{i)} the relative right ascension ($\alpha$) and \textit{ii)} declination ($\delta$) of the star, namely the angular separations between the central object and the orbiting star a different epochs, projected on the observer's sky plane, and \textit{iii)} its LOS velocity ($v_{\rm LOS}$). 
In order to do this in a fully relativistic way, null photon paths should be integrated from the point of emission up to the distant observer. This, in turn, requires solving the emitter-observer problem to identify the connecting photon an then computing the photon travel time, to obtain the time of emission and the projected sky-position. A numerical methodology for performing both operations has been introduced in \citep{DellaMonica2023b}. Once this is applied we can reconstruct fully relativistic astrometric observables for the S2 stars. In order to assess the necessity to apply this costly procedure for the S2 star, we have preliminarily considered a general relativistic orbit for this star (obtained by setting $\gamma=1$, being the only parameter that affects the photon propagation) and than applied the photon-path reconstruction procedure for different values of $\gamma$ to assess the astrometric impact. For any given observation time, $t_\textrm{obs}$, we can compute astrometric observables for some values of $\gamma$, ($\alpha_{\gamma}(t_\textrm{obs})$, $\delta_{\gamma}(t_\textrm{obs})$) and for the general relativistic case, ($\alpha_{\gamma=1}(t_\textrm{obs})$, $\delta_{\gamma=1}(t_\textrm{obs})$). The two can then be compared by computing the astrometric deviation
\begin{equation}
    \Delta_A = \sqrt{(\alpha_{\gamma}-\alpha_{\gamma=1})^2+(\delta_{\gamma}-\delta_{\gamma=1})^2}.
    \label{eq:astrometric_deviation}
\end{equation}
The results of this comparison are shown in Fig. \ref{fig:astrometric_lensing} for values of $\gamma\in[1,\,2]$ for epochs around the pericenter passage where relativistic effects on the emitted photons are supposed to be larger. Our analysis shows that for the S2 star, even a value of $\gamma$ that deviates by 100\% from the general relativistic values would produce a change in the astrometric observable, only due to the different photon path propagation, that is 
below the nominal sensitivity threshold for GRAVITY observations 
which is fixed at $\sim10\,\mu$as. 
Therefore, we conclude that effects of photon propagation related to a value of $\gamma\neq 1$ are negligible. This guarantees us that we can estimate the astrometric observables for the S2 without having to resort to the null-path integration, which is replaced by the much computationally cheaper geometric projection of the star's trajectory. The transformation from the black hole reference frame to that of a distant observer is achieved through a rotation by the angles $i$, $\Omega$, and $\omega$ (encoded in Thiele-Innes elements), followed by a translation by the distance $D$ from the Galactic Center  to the Earth-based observatory along the 
LOS. This transformation also includes the calculation of the classical Rømer effect, which is the only photon delay effect that produces non-negligible shifts in the received signal (see also the supplementary materials of \citep{Do2019b}).

Special attention is required when computing the LOS velocity, because of the presence of relativistic effects. First, the star’s kinematic velocity is projected along the LOS. This projected velocity is then converted into the classical longitudinal Doppler shift, which accounts for most of the observed frequency shift. Additionally, 1PN relativistic effects contribute to the redshift. These include special and general relativistic time dilation effects, which are significant due to S2’s high kinematic velocity (approximately $7700$ km/s) and the gravitational potential at pericenter ($\Phi/c^2 = GM/rc^2 \sim 3 \times 10^{-4}$). These relativistic effects introduce an additional redshift ($\sim200$ km/s at the pericenter) that has been experimentally confirmed \citep{Do2019b, GravityCollaboration2018a}. The extra redshift contribution is derived from the time component of the integrated 4-velocity, $\dot{t} = dt/d\tau$ and does indeed depend on $\gamma$ only, potentially allowing for the possibility of breaking the degeneracy between the PPN parameters.

\begin{figure}
    \centering
    \includegraphics[width=\columnwidth]{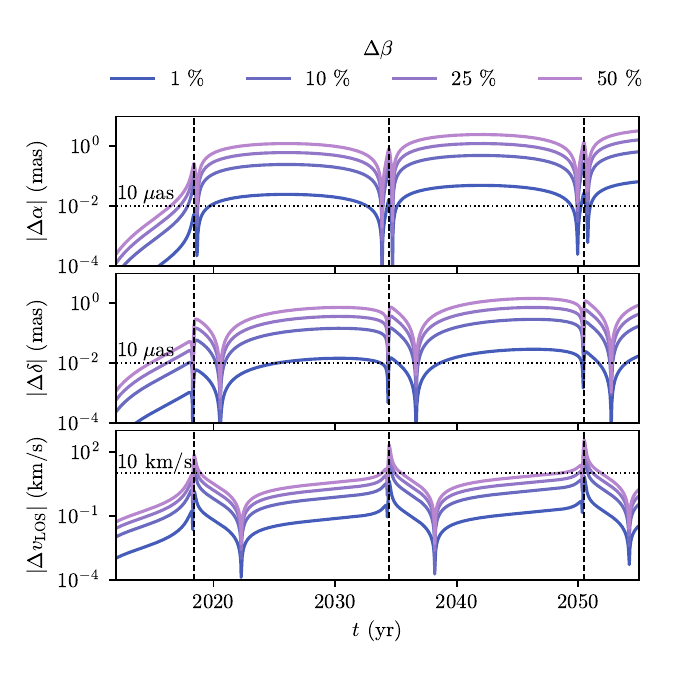}
    \caption{Astrometric (right ascension $\alpha$ and declination $\delta$, top two panels) and spectroscopic (LOS velocity, bottom panel) impact of considering a PPN parameter $\beta\neq 1$, for different values of percentage deviation from unity, for the S2 star. The horizontal dotted lines correspond to the nominal accuracy of the GRAVITY interferometer and the SINFONI spectrographer, for astrometric and spectroscopic observables respectively.}
    \label{fig:astrometric_beta}
\end{figure}

\begin{figure}
    \centering
    \includegraphics[width=\columnwidth]{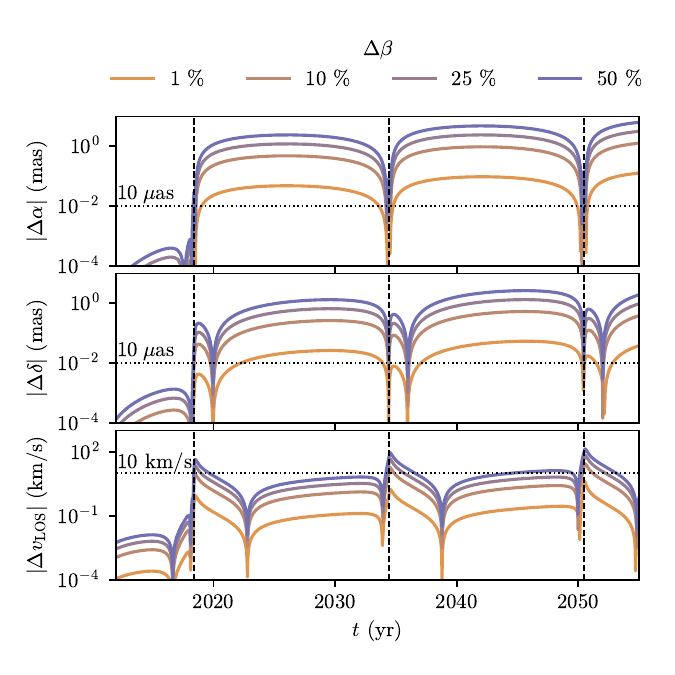}
    \caption{As in Fig. \ref{fig:astrometric_beta} but for the parameter $\gamma$.}
    \label{fig:astrometric_gamma}
\end{figure}

Using our orbital model we can quantitatively estimate the impact on the observables of changing the PPN parameters with respect to their relativistic value. In Figs. \ref{fig:astrometric_beta} and \ref{fig:astrometric_gamma} we display the astrometric and spectroscopic deviations from the $\gamma=\beta=1$ 
case for the S2 star, computed by considering values of $\beta$ and $\gamma$ that differ from one by some percentage deviation, and by only changing one of them at a time, while considering the other fixed to unity. 
We compare the deviation in the astrometric and spectroscopic observables with the nominal accuracy of the GRAVITY interferometer and the SINFONI spectrographer, 
respectively, 
and we 
find that a variation as small as 1\% on both $\beta$ and $\gamma$ is able to bring the astrometric deviation above the sensitivity threshold for GRAVITY. The same does not apply to the spectroscopic observable, for which the deviations exceed the nominal sensitivity threshold for SINFONI only for much larger departures from the unity ($\sim10-25\%$) of the PPN parameters. These results qualitatively suggest that using orbital data for the S2 star in the Galactic Center will only be able to put a constraints on the combination $\Upsilon$ of PPN parameters that regulates the orbital dynamics, \textit{i.e.} the one entering in Eq. \eqref{eq:PPN_precession}, and not on the two parameters separately, due to the much smaller impact of the photon propagation effects (in this case the gravitational redshift considered in the reconstruction of the 
LOS velocity) on the observables. In order to assess this more quantitatively, in the next section we will perform a full posterior analysis for our orbital model.

\section{MCMC analysis}
\label{sec:mcmc}

In this section we explore whether future data from the GRAVITY interferometer, particularly for the S2 star, can constrain and to which extent, the PPN parameters in the space-time presented in Eq. \eqref{eq:1PPN_metric}. To conduct this forecast analysis, we use the orbital model developed in the previous section to 
fit 
a mock catalog of future observations that realistically replicate the sensitivity and limitations of the VLT instruments GRAVITY (for astrometry) and SINFONI (for spectroscopic measurements of LOS velocities) over two entire orbital periods (thus covering at least one pericenter passage and full subsequent orbit), starting from the most recent pericenter passage in $\sim 2018$. 
The mock catalog was built 
using, as fiducial model, the Schwarzschild metric, whose 1PN approximation coincide with Equations \eqref{eq:1PPN_metric}-\eqref{eq:grr} where $\gamma=\beta=1$, and was first introduced in  \citet{DellaMonica2022b} and can be accessed publicly\footnote{Available on \href{https://github.com/rdellamonica/S2-gr-mock-catalogue}{Github}.}. In our 
fiducial model, we also assume: \textit{i)} that the \textit{true} values of the orbital parameters of S2 correspond to the best-fit values from \citep{GravityCollaboration2020}; \textit{ii)} that astrometric and spectroscopic observations occur at the same epochs, following an observational cadence that becomes denser around the pericenter passage. Specifically, following \citep{Grould2017}, we assume the following observational strategy:
\begin{itemize}
    \item one observation per day during the two weeks centered on the pericenter passage;
    \item one observation every two nights over the month centered on the pericenter passage;
    \item one observation per week during the two months centered on the pericenter;
    \item one observation per month during the rest of the year of the pericenter passage;
    \item two observations per year in the remaining years.
\end{itemize}
We thus construct an array of observation times, $\{t_{\textrm{obs}}\}$, and use this to compute the astrometric and spectroscopic observables. Additionally, \textit{iii)} we assume the instruments operate under ideal conditions, acquiring data at their maximum nominal accuracies: $\sigma_A = 10\,\mu\textrm{as}$ for GRAVITY astrometry, and $\sigma_V = 10\,\textrm{km/s}$ for SINFONI LOS velocities. These uncertainties are accounted for by adding Gaussian noise with zero mean and standard deviations $\sigma_A$ and $\sigma_V$ to the astrometric data and LOS velocities, respectively.

Since our baseline model is built upon the assumption of the Schwarzschild metric, 
and because of the validity of the weak-field limit approximation for S2, in our posterior analysis we will estimate the ability to recover the parameters $\gamma=\beta=1$, corresponding to the general relativistic case. The qualitative analysis from the previous section showed that variations of order 1\% on these PPN parameters are expected to produce detectable deviations in the astrometric observables. Hence, we use these heuristic limits on $\beta$ and $\gamma$ as uniform priors for our analysis. Moreover, for the orbital parameters we set large uniform priors centered around the values used for the generation of the mock catalog. The full set of uniform priors is reported in Table \ref{tab:priors_posteriors}. We compute the likelihood of a given set of parameters in Eq. \eqref{eq:parameters} using the following function
\begin{align}
    \log \mathcal{L} = -\frac{1}{2}\sum_i^{N}\biggl[&\biggl(\frac{\alpha_i(\vec{\theta})-\alpha^{{mock}}_{i}}{\sigma_A}\biggr)^2+ \nonumber\\
    +&\biggl(\frac{\delta_i(\vec{\theta})-\delta^{{mock}}_{i}}{\sigma_A}\biggr)^2+\label{eq:likelihood_mock}\\
    +&\biggl(\frac{v_{\textrm{LOS}, i}(\vec{\theta})-v_{\textrm{LOS}, i}^{mock}}{\sigma_V}\biggr)^2\biggr],\nonumber
\end{align}
where quantities without a superscript correspond to the observables reconstructed for the given set of parameters $\vec{\theta}$, while quantities with the $mock$ superscript refers to the data from our mock catalog. The uncertainties $\sigma_A = 10\,\mu$as and $\sigma_V=10$ km/s are the nominal astrometric and spectroscopic uncertainties from the GRAVITY interferometer and the SINFONI spectrographer, respectively, used in the generation of the mock catalogue itself. For this study, we focus on two full orbital periods which, combined with the observational strategy highlighted in the previous paragraph correspond to a total number of $N=212$ epochs of observation.

Given the aforementioned priors and the likelihood defined in Eq. \eqref{eq:likelihood_mock} we draw samples from the posterior distribution using an affine-invariant MCMC sampler implemented in \citep{ForemanMackey2013}, using a convergence check based on the estimation of the mean auto-correlation time for the walkers. The results of our analysis are presented in the next section.

\begin{table*}[]
    \centering
    \setlength{\tabcolsep}{16pt}
    \renewcommand{\arraystretch}{1.7}
    \caption{Priors and results of the MCMC analysis.}
    \begin{tabular}{|lccc|}
        \hline
        Parameter (unit) &   \multicolumn{2}{c}{Uniform priors}& Posterior \\ \cline{2-3}
        &  Start & End &\\
        \hline
        $D$ (kpc) &   8.24&8.25&$8.2474\pm0.0024$ \\
        $M$ ($10^6M_\odot$) &   4.25&4.27&$4.2620\pm0.0037$ \\
        $t_p$ (yr) &   2018.3789&2018.3791&$2018.378990_{-0.000022}^{+0.000021}$ \\
        $a$ (arcsec) &   0.1250&0.1251&$0.1250600_{-0.0000028}^{+0.0000029}$ \\
        $e$ &   0.884&0.885&$0.8846535_{-0.0000050}^{+0.0000049}$ \\
        $i$ ($^\circ$) &   134.5&134.6&$134.5651_{-0.0017}^{+0.0016}$ \\
        $\Omega$ ($^\circ$) &   228.1&228.2&$228.1710\pm0.0025$ \\
        $\omega$ ($^\circ$) &   66.2&66.3&$66.2655\pm0.0024$ \\ \hline
        $\beta$& 0.9& 1.1& \multirow{2}{*}{$\Upsilon = 1.000\pm0.004$}\\
        $\gamma$& 0.9& 1.1&\\ \hline
    \end{tabular}
    \tablefoot{The table displays the full set of uniform priors used for our MCMC analysis (columns two and three) and the resulting posteriors ($1\sigma$ confidence interval) for all the parameters in our model. The parameters $\beta$ and $\gamma$ could not be constrained separately from our analysis, but their combination $\Upsilon$ is well constrained with a precision of $\sim 0.4\%$.}
    \label{tab:priors_posteriors}
\end{table*}

\section{Results for S2}
\label{sec:results}

\begin{figure}
    \centering
    \includegraphics[width=\columnwidth]{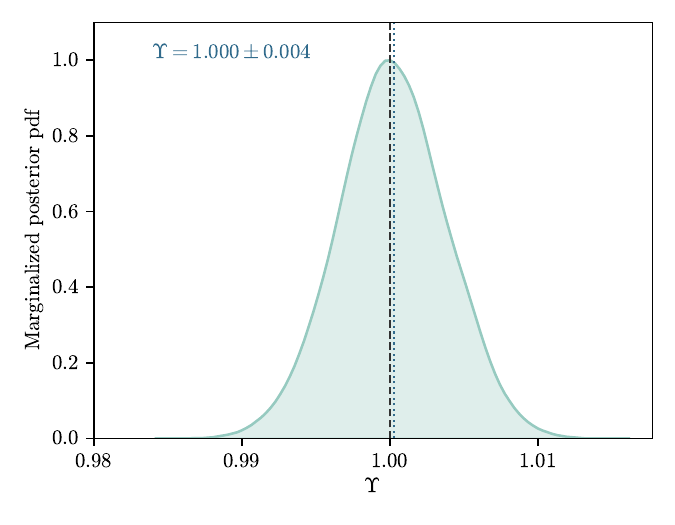}
    \caption{Marginalized posterior probability density function on the parameter $\Upsilon$ in Eq. \ref{eq:upsilon} derived from the samples of our full posterior distribution. The posterior is perfectly centered on the general relativistic value, $\Upsilon = 1$, used as input for the generation of the mock catalog and reaches a precision of order $0.4\%$ at 1$\sigma$.}
    \label{fig:upsilon_posterior}
\end{figure}

In this section, we present the results of our forecast analysis for the S2 star. In the right-most column of Table \ref{tab:priors_posteriors}, we report the median-centered 68\% confidence intervals on the orbital parameters of our model retrieved by the sampled posterior distribution. This $1\sigma$ bounds perfectly recover the input values of the parameters for the generation of the mock catalog. As expected, our posterior analysis is not able to put a limit on either PPN parameter $\beta$ or $\gamma$, due to the impossibility using current orbital data, only marginally affected by relativistic photon-effects, to break the degeneracy between the two parameters. Nonetheless, when marginalizing over all the orbital parameters, and deriving the posterior probability distribution on the combination $\Upsilon$ of PPN parameters, in Eq. \eqref{eq:upsilon}, which we report for completeness in Fig. \ref{fig:upsilon_posterior}, we obtain a value of $\Upsilon$ perfectly centered on $\Upsilon=1$, with an uncertainty of order $0.4\%$. Importantly, we have repeated the analysis by considering larger priors on the PPN parameters, namely $\beta\in[0,\,2]$ and $\gamma\in[0,\,2]$, and obtained posteriors that match both qualitatively and quantitatively with the results of the analysis that we have just presented. This confirms that the inability of setting a constraint on $\gamma$ using S2 data is not a spurious effect of the narrow priors used.

\section{The potential lensing measurement for S62}
\label{sec:s62}

\begin{figure}
    \centering
    \includegraphics[width=\columnwidth]{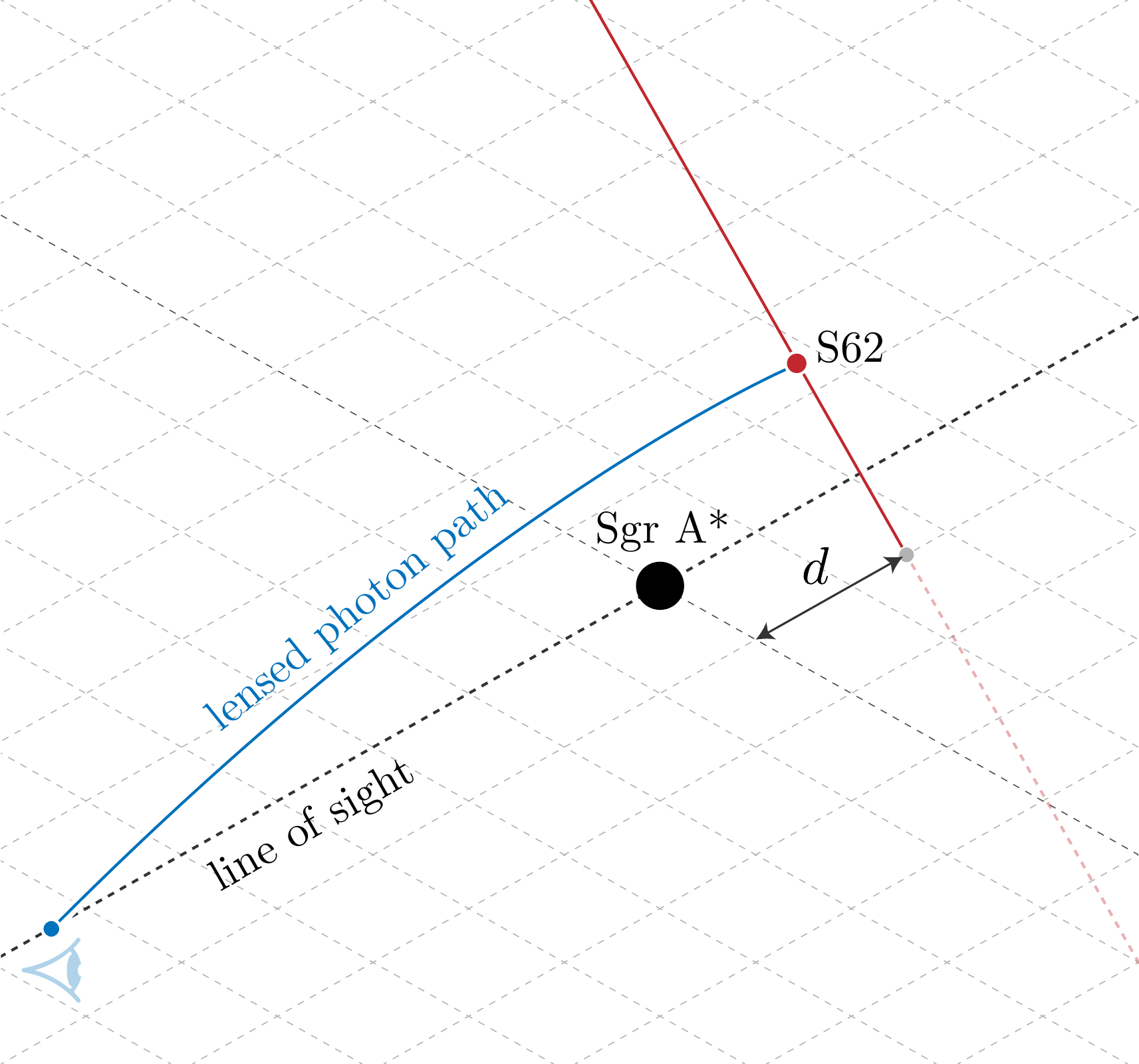}
    \caption{Geometrical configuration of S62. We consider the star to move on a plane perpendicular to the LOS, placed at an unknown distance $d$ from Sgr A*. The interesting lensing effects occur when this plane lies behind Sgr A* with respect to an Earth-based observer.}
    \label{fig:s62}
\end{figure}

\begin{figure}
    \centering
    \includegraphics[width=\columnwidth]{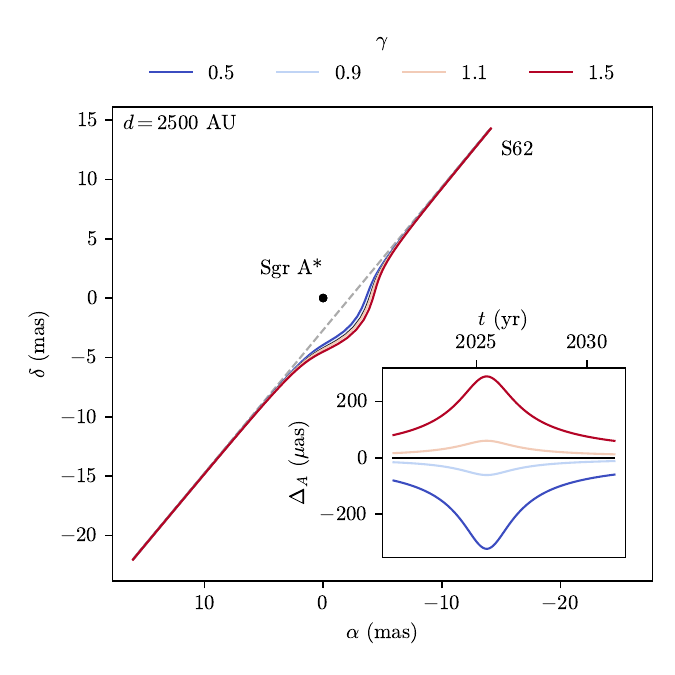}
    \caption{Reconstructed sky-positions of S62 for an example value of $d = 2500$ AU of the plane on which the trajectory of this star lies (see Fig. \ref{fig:s62}). The gray dashed line represent the unlensed sky-position of S62, while solid colored lines correspond to different values of $\gamma$. The inset plots displays as a function of time the astrometric deviation in $\mu$as between the cases with $\gamma \neq 1$ and the general relativistic case with $\gamma = 1$. Depending on the value of $\gamma$, this deviation can surpass the sensitivity threshold for GRAVITY, implying the ability to potentially constrain this PPN parameter using only astrometric observations.}
    \label{fig:s62-lensing}
\end{figure}

With our previous analysis, we have shown that considering the orbital data for S2 star is not sufficient to break the degeneracy between $\beta$ and $\gamma$, due to the small amplitude of photon-related relativistic effects in the astronomic observables of this star. In this section we analyze the possibility to break this degeneracy by a direct measurement of an independent lensing event for a star that pass much closer in the sky to Sgr A*, which would enable the individual measurement of the PPN parameter $\gamma$ (which, alone, regulates the amplitudes of relativistic effects on the photons). Deep images of the Galactic Center with GRAVITY \citep{GravityCollaboration2022} obtained between 2019 and 2021 have revealed the presence of a slowly moving star, S62, faintly visible in the K-band, which is slowly approaching Sgr A*. The sky-projected motion of S62 over the different observation epochs has not shown any sign of orbital acceleration, so that the current trajectory of this object has been modeled as a linear motion
\begin{align}
    \alpha(t) &= \alpha_0 + v_\alpha (t-t_0)\label{eq:linear_trajectory_1}\\
    \delta(t) &= \delta_0 + v_\delta (t-t_0)\label{eq:linear_trajectory_2}
\end{align}
with S62 being at $\alpha_0 = -14.1$ mas and $\delta_0 = 13.6$ mas at $t_0 =$ 2021-03-01, and the linear drift given by $v_\alpha = 2.97$ mas yr$^{-1}$ and $v_\delta = -3.58$ mas yr$^{-1}$. For simplicity, we consider the plane on which S62 moves linearly to be perpendicular to LOS and placed at an unknown distance $d$ from Sgr A*. In order for lensing effects to be observable by a distant observer (that we consider placed at an Earth-based observatory, \textit{i.e.} at a distance of $D = 8$ kpc) the plane on which the star moves has to be behind Sgr A* with respect to the observer (see the geometrical configuration of the system we are considering in Fig. \ref{fig:s62}). It is important to remark that, to the date, the actual three-dimensional trajectory of S62 is unknown and its orbital properties are currently object of debate \citep{Peissker2020, GravityCollaboration2022, Peissker2022}, with a recent re-analysis of two-decades worth of NACO observations pointing towards a very-tight $\sim$10 years orbit for S62 \citep{Peissker2020}. We will nonetheless use our working assumption to show, as a proof-of-concept, that a single detection of a lensing event for this star can help constrain the value of the PPN parameter $\gamma$ independently from $\beta$. In order to do so, we consider the fully-relativistic astrometric observables produced by integrating null photon paths from the putative trajectory of S62 up to the distant observer, thus including the lensing effect and the Shapiro delay in the reconstructed sky positions. To obtain this, we apply the same procedure used for the fully-relativistic reconstruction of observables of the S2 star, based on the methodology introduced in \citep{DellaMonica2023b}. For each position of the test particle on its linear trajectory in Eqs. \eqref{eq:linear_trajectory_1}-\eqref{eq:linear_trajectory_2}, our procedure returns the plane on which the photon propagates (identified by a position angle of the plane with respect to the chosen reference direction), an impact parameter $b$ of the photon, uniquely identifying a photon on such plane and a relativistic propagation time. From the impact parameter and the position angle of the plane on which the photon propagates, one can construct relativistic astrometric observables, ($\alpha_\textrm{lensed},\,\delta_\textrm{lensed}$), while from the relativistic propagation time, we reconstruct the coordinate time of observation $t_\textrm{obs}$. The resulting sky-projected lensed position for S62 for an example distance $d=2500$ AU are shown in Fig. \ref{fig:s62-lensing} for different values of the PPN parameters $\gamma$. In the inset plot, we also show the astrometric deviation, computed as in Eq. \eqref{eq:astrometric_deviation}, between the general relativistic case ($\gamma=1$) and those with $\gamma \neq 1$. These deviations are symmetric for values of $\gamma$ that have a given deviation from unity, and correspond to a larger amount of the the gravitational lensing for values of $\gamma > 1$ and a less prominent lensing for $\gamma < 1$. Moreover, depending on the value of $\gamma$, this deviation can surpass the sensitivity threshold for GRAVITY, which implies the potential ability to constrain the PPN parameter $\gamma$ using only astrometric observations of lensed stars at the Galactic Center. 

\begin{figure*}
    \includegraphics[width=\textwidth]{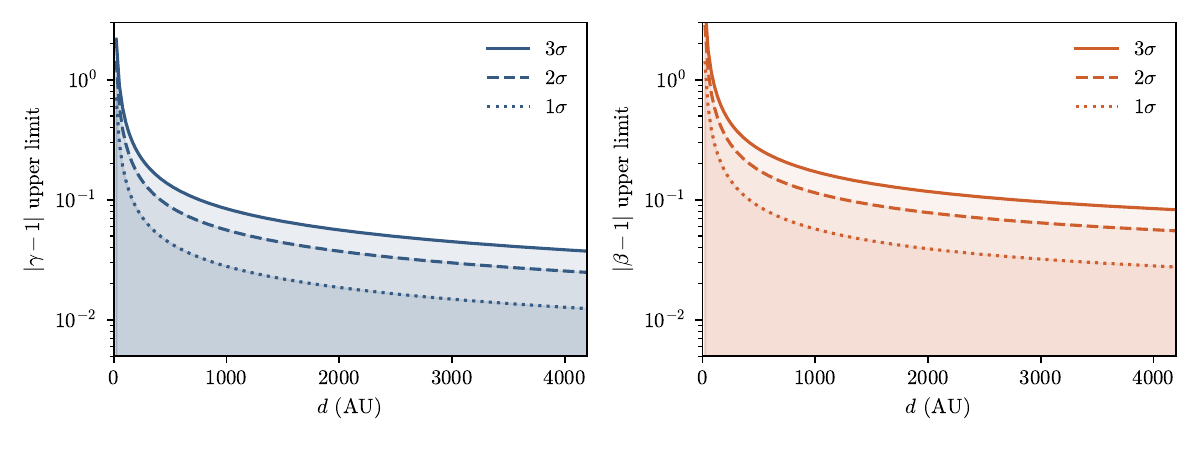}
    \caption{\textit{Left panel:} upper limit on $|\gamma-1|$ at $1\sigma$, $2\sigma$ and $3\sigma$ as a function of the distance $d$ of S62 plane, perpendicular to the LOS, from Sgr A*. When the S62 plane crosses Sgr A*, the upper limit on $\gamma$ tends to diverge. For values of $d > 0$, on the other hand, $|\gamma-1|$ is bound an the limit reaches $|\gamma-1|\lesssim 1.2\times 10^{-2}$ (at $1\sigma$) in correspondence with $d\approx 4000$ AU. \textit{Right panel:} upper limit on $|1-\beta|$ at $1\sigma$, $2\sigma$ and $3\sigma$ as a function $d$, obtained by combining the forecasted constraints on the combination $\Upsilon$ of PPN parameters, obtained from our analysis on the S2 star, and the lensing constraints on $\gamma$ forecasted for S2.}
    \label{fig:s62-lensing-gamma-beta}
\end{figure*}

From our determination of the astrometric deviations as a function of the PPN parameter $\gamma$, we can directly estimate for different values of the distance $d$ of the orbital plane of S62 from Sgr A* what would be the departure from unity of $\gamma$ that would produce a deviation measurable at $1\sigma$, $2\sigma$ and $3\sigma$ with the nominal GRAVITY sensitivity. In order to do so, we consider values of $d\in[0, 4000]$ AU and for each of them we solve for the value of $\gamma$ that produces $\Delta_A = 10\,\mu$as, $20\,\,\mu$as and $30\,\mu$as (thus assuming a nominal accuracy of GRAVITY of $10\,\mu$as) when this quantity is maximized, namely in the point of closest approach between S62 and Sgr A* on the sky-plane. In this case, we consider only modifications to the photon paths, thus leaving the actual trajectory of S62 unaltered and only considering the difference in the reconstructed observables for different values of $\gamma$. We interpret the values of $\gamma$ obtained with this approach as an upper limit on $|\gamma-1|$ at $1\sigma$, $2\sigma$ and $3\sigma$, respectively and we report the results of our analysis in the left panel of Fig. \ref{fig:s62-lensing-gamma-beta}. As expected, when the distance from the S62 plane and Sgr A* tends to 0, this upper limit tends to diverge, as the lens effect produced by Sgr A* (without considering any other orbital effect) is canceled when the star plane cross it. However, for values of $d > 0$, the upper limit of $|\gamma-1|$ decreases and reaches $|\gamma-1|\lesssim 1.2\times 10^{-2}$ in correspondence with $d\approx 4000$ AU. An important caveat to take into consideration in this analysis is that the numbers that we obtain for the upper limits on $|\gamma-1|$ would correspond to what one would actually obtain, only in the case of having obtained by an independent measurement the distance $d$ of the S62 plane and any other quantity that characterizes the trajectory of S62 on this plane. If, on the other hand, $d$ is obtained through the same orbital data that are included in the determination of the upper limit $|\gamma-1|$, then the degeneracy between these two parameters must be taken into account which might bring to a degradation of the constraints that we have shown. 

Finally, from the combination of the forecasted constraints obtained on $\gamma$ for the S62 star, and those previously obtained on $\Upsilon$ from S2, shown in Fig. \ref{fig:upsilon_posterior}, we can estimate the corresponding constraints obtained on $\beta$, that are shown in the right panel of Fig. \ref{fig:s62-lensing-gamma-beta} as a function of the unknown distance $d$. Similarly, to the case of $\gamma$, the greater the distance $d$, the smaller the confidence interval around unity of $\beta$ is obtained, with a precision of $\sim2\%$, corresponding to $|\gamma-1|\lesssim 2.4\times 10^{-2}$, for $d\approx 4000$ AU. As $d$ approaches 0, the interval widens due to the increasingly smaller amplitude of the lensing effects.

\section{Discussion and conclusions}
\label{sec:discussion}

The PPN formalism offers a comprehensive and agnostic framework for testing theories of gravity beyond General Relativity \citep{Will1972}. By introducing a set of dimensionless parameters, such as $\gamma$ and $\beta$, the formalism systematically quantifies deviations from General Relativity and enables consistent comparisons between different theories of gravity \citep{Psaltis2008}. It provides a valuable tool for exploring both weak- and strong-field regimes, making it particularly relevant for astrophysical tests in the Solar System \citep{Will2014} and around massive objects like the supermassive black hole at the Galactic Center \citep{DeLaurentis2023}.

In this work, we have shown that tracking the orbital motion of stars such as S2 with the currently available most precise near-infrared instrument for observations at the Galactic Center, \emph{i.e.} GRAVITY, can potentially provide important constraints on the PPN parameters. However, since photon-lensing effects are sub-dominant for this source, such observations alone can only constrain a combination of $\gamma$ and $\beta$. Specifically, by fitting a PPN orbital model to a mock catalog of data based on the Schwarzschild space-time (thus assuming the validity of General Relativity in our fiducial model) covering two full periods of S2, we have forecasted to recover a posteriori the linear combination $\Upsilon$ of PPN parameters, introduced in Eq. \eqref{eq:upsilon}, with a tight constraint $\Upsilon = 1.000 \pm 0.004$. Moreover, the introduction of potential astrometric measurements of lensing effects (that solely depend on the value of $\gamma$) in the sky-projected trajectory of the slowly moving star S62, when combined with the orbital tracking of S2, can deliver independent constraints on both parameters, with potential precisions as fine as 1\% for $\gamma$ and 2\% for $\beta$. Such results are based on several assumptions that should be acknowledged as potential sources of systematics and underestimation of errors. This is especially relevant for what concerns the three-dimensional motion of the S62 star. Since current observational data do not fully constrain the star's trajectory \citep{GravityCollaboration2022} and the ongoing controversy on the actual geometrical configuration of its orbit \cite{Peissker2020, GravityCollaboration2022, Peissker2022}, our estimates of lensing effects involve some degree of uncertainty and depend on the unknown geometrical configuration of S62's orbit and its distance $d$ from Sgr A*.  Future tracking of this object in the upcoming years will be crucial to refine these estimates and obtain robust constraints on the PPN parameters.

While our forecasted accuracy for $\beta$ and $\gamma$ hints at the potential for interesting tests of gravity in a regime that was previously unexplored for this formalism, the precision obtained in the Solar Systems \citep{Bertotti2003, Genova2018} remains unmatched. The next generation of infrared telescopes, such as the proposed improvement of VLT facilities, GRAVITY+ \citep{GravityPlus2022}, or MICADO \citep{Sturm2024} on the Extremely Large Telescope (ELT) and the Thirty Meter Telescope in Hawaii \citep{Skidmore2015}, will significantly expand our ability to observe the Galactic Center with a potential improvement of the constraints that we have forecasted here. In fact, these instruments, with their superior sensitivity and resolution, will enable not only more precise tracking of stellar orbits and detection of relativistic effects but also the discovery of fainter stars on tighter orbits around Sgr A* \citep{GravityCollaboration2022}.

Additionally, an independent measurement of strong gravitational lensing by the SMBH in the Galactic Center is provided by the observation of the black hole shadow for Sgr A* by the Event Horizon Telescope \citep[EHT]{EventHorizonTelescopeCollaboration2022a}, with the potential to provide an independent constraint on $\gamma$ and $\beta$, breaking the degeneracy from S-stars observations. However, with the current resolution, the EHT can only track the diameter of the radio luminosity peak of the accretion flow (which serves as a proxy for an estimation of the geometrically-relevant photon ring and the black hole shadow). This, united with the fact that EHT analyses of the fractional deviation from General Relativity of the observed shadow diameter rely on priors on the mass-over-distance ratio that are derived from S-stars observations \citep{EventHorizonTelescopeCollaboration2022b}, makes it difficult to quantitatively convert the currently measured diameter of the shadow into constraints on the 1PN parameters.

Nonetheless, a major breakthrough in the characterization of the gravitational field of the Galactic Center and in the precision of gravity tests therein is expected with the discovery of potential pulsars at the Galactic Center \citep{DellaMonica2023b, DellaMonica2025}. With their stable timing signatures, pulsars can provide highly accurate measurements of relativistic time delays and other effects, making them ideal probes for testing deviations from General Relativity and thus one of the main scientific goals of future facilities like the Squared Kilometer Array \citep{Weltman2020}.

\begin{acknowledgements}
IDM and RDM acknowledge support from the  grant PID2021-122938NB-I00 funded by MCIN/AEI/10.13039/501100011033 and by ``ERDF A way of making Europe''.  IDM also acknowledges financial support from the grant SA097P24 funded by Junta de Castilla y León. RDM also acknowledges support from the Consejeria de Educación de la Junta de Castilla y León and the European Social Fund. 
\end{acknowledgements}

\bibliographystyle{aa} 
\bibliography{biblio}

\end{document}